%% file: DRAFT_sigcomm16.tex
\documentclass[10pt]{sig-alternate-05-2015}

\usepackage{array}
\usepackage{graphicx}
\usepackage{float}
\usepackage{stfloats}
\usepackage{lipsum}
\usepackage{tabularx}
\usepackage[tight,footnotesize]{subfigure}
\usepackage{caption}
\usepackage{epstopdf}
\usepackage{amsmath}
\usepackage{amssymb}
\usepackage{paralist}
\usepackage{mathrsfs}
\usepackage{url}
\usepackage{mathbbol}
\usepackage{times}

\usepackage{booktabs}
\usepackage{verbatim}

\usepackage{bm}			
\usepackage{cite}

\usepackage{amsthm}

\usepackage[table]{xcolor}
\usepackage{color,colortbl}

\definecolor{Gray}{gray}{0.9}

\usepackage[normalem]{ulem}

\begin{document}


%

\CopyrightYear{2016}
\setcopyright{none}
\conferenceinfo{SIGCOMM '16,}{August 22--26, 2016, Florianopolis, Brazil}
\isbn{978-1-4503-4193-6/16/08}
\doi{http://dx.doi.org/10.1145/2934872.2959078}

%
%
%

%


\title{ARTEMIS: Real-Time Detection and Automatic Mitigation for BGP Prefix Hijacking}

\author{
Gavriil Chaviaras, Petros Gigis, Pavlos Sermpezis, and Xenofontas Dimitropoulos\\
\affaddr{FORTH / University of Crete,  Greece}\\
\email{\{gchaviaras, gkigkis, sermpezis, fontas\}@ics.forth.gr}
}

\maketitle

\begin{abstract}
\input{Abstract}
\end{abstract}
%
%

\vspace{-1.5\baselineskip}



\vspace{-0.25\baselineskip}


\section{Introduction}\label{sec:intro}
\input{Introduction}

\section{ARTEMIS Overview}\label{sec:overview}
\input{System}

\section{Experiments with a real AS}\label{sec:experiments}
\input{Experiments}

\section{Demo}\label{sec:demo}
\input{Demo}

\noindent\textbf{Acknowledgements.} This work has been funded by the European Research Council Grant Agreement no. 338402.
\vspace{-0.5\baselineskip}

\bibliographystyle{ieeetr}


\end{document}

%% file: Abstract.tex
Prefix hijacking is a common phenomenon in the Internet that often causes routing problems and economic losses. In this demo, we propose ARTEMIS, a tool that enables network administrators to \textit{detect and mitigate} prefix hijacking incidents, against their own prefixes. ARTEMIS is based on the real-time monitoring of BGP data in the Internet, and software-defined networking (SDN) principles, and can completely mitigate a prefix hijacking within a few minutes (e.g., 5-6mins in our experiments) after it has been~launched.

%% file: Introduction.tex
The Internet is composed of thousands of Autonomous Systems (ASes), whose inter-domain traffic is routed with the Border Gateway Protocol (BGP). Due to the distributed nature and lack of authorization in BGP, an AS can advertise illegitimate paths or prefixes owned by other ASes, i.e., hijacking their prefixes. Prefix hijacking can cause serious routing problems and economic losses. For instance, YouTube's prefixes were hijacked in 2008 disrupting its services for more than $2$ hours~\cite{hijack-YouTube}, whereas China Telecom hijacked $37 000$ prefixes (about $10\%$ of the BGP table) in 2010 causing routing problems in the whole Internet for several minutes~\cite{hijack-ChinaTelecom}. 

Prefix hijacking (due to an attack or misconfiguration) is a common phenomenon in the Internet, and since its prevention is not always possible, mechanisms for its detection and mitigation are needed. To this end, several methodologies for detecting prefix hijackings have been proposed, e.g., ~\cite{Shi-Argus-IMC-2012,Lad-Phas-Usenix-2006}. However, most previous works focus on \textit{alert systems} that are not controlled by the AS itself~\cite{Shi-Argus-IMC-2012,Lad-Phas-Usenix-2006}, but offer BGP prefix hijacking detection as a service to ASes. 
In addition, previous research focuses primarily on {\it accurately} detecting BGP hijacks, rather than {\it timely} detecting {\it and} mitigating them. The whole detection/mitigation cycle presently has significant delay: (i) aggregated BGP data from RouteViews~\cite{routeviews} or RIPE RIS~\cite{ripe-ris}, which are commonly-used for detection, become available approximately every $2$ hours (BGP full RIBs) or $15$mins (BGP updates)
; (ii) a network administrator that receives a notification from a third-party alert system needs to {\it manually} process it to verify if the notification corresponds to a hijacking or is a false alarm; and (iii) for mitigation, administrators often need to manually reconfigure routers or contact administrators of other ASes to filter announcements. YouTube, for example, reacted about 80min after the hijacking of its prefixes. These problems render existing 
mechanisms inefficient especially for a large percentage of hijacking events that last only for a short time (cf., more than $20\%$ of hijacks last $<10$mins~\cite{Shi-Argus-IMC-2012}).

In this work, our goal is to enable network administrators to {\it timely} detect and mitigate prefix hijacking incidents, e.g., in 5-6 mins, against their {\it own} prefixes. To accelerate detection, our approach exploits real-time BGP data from: (i) Looking Glass (LG) servers; and (ii) BGP collectors with live data streaming capabilities, which are provided by the RIPE RIS~\cite{ripe-ris,ripe-ris-real-time} and BGPmon~\cite{bgpmon} projects. LGs provide a view directly from operational BGP routers, without intermediate collectors, while (the recent) RIPE RIS streaming service~\cite{ripe-ris-real-time} and BGPmon~\cite{bgpmon} provide real-time feeds of the collected BGP data. Furthermore, we {\it automatically} mitigate hijackings of prefixes owned by an AS by announcing de-aggregated BGP prefixes. We combine these in a tool we call ARTEMIS (\textit{Automatic and Real-Time dEtection and MItigation System}), which can detect a prefix hijack in near real-time, and mitigate it without any manual intervention.

We evaluated ARTEMIS in real settings, by deploying it to detect and mitigate prefix hijackings performed against our own prefixes from an actual AS in the Internet. We found that we can detect hijacks in <$1$min, start the mitigation in a few seconds, and completely solve the problem in around $5$mins. To our best knowledge, this is the first time that we can \textit{detect and mitigate} hijacks within a few minutes. 

%% file: System.tex
\begin{figure}
\centering
\includegraphics[width=0.8\linewidth]{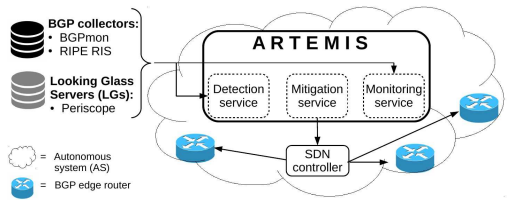}
\caption{ARTEMIS overview.}
\label{fig:architecture}
\vspace{-1\baselineskip}
\end{figure}

ARTEMIS consists of three components: a \textit{detection}, a \textit{mitigation}, and a \textit{monitoring} service as shown in Fig.~\ref{fig:architecture}. The detection service runs continuously and combines \textit{control plane} information from Periscope~\cite{Giotsas-Periscope-PAM-2016} (an LG API),  the streaming service of RIPE RIS~\cite{ripe-ris-real-time}, and BGPmon~\cite{bgpmon}, which return in near real-time BGP routes/updates for a given list of prefixes and ASNs
. 
By combining multiple sources, the delay of the detection phase is the min of the delays of these sources. The system can be parametrized (e.g., selecting LGs based on location or connectivity) to achieve trade-offs between monitoring overhead and detection efficiency/speed.

When a prefix hijacking is detected, ARTEMIS launches the mitigation service, which changes the configuration of BGP routers to announce the de-aggregated sub-prefixes of the hijacked prefix. Therefore, ARTEMIS assumes permissions for sending BGP advertisements for the owned prefixes from the BGP routers of the network. This can be effectively accomplished by running ARTEMIS, as an application-level module, over a network controller that supports BGP, like ONOS~\cite{onos} or OpenDayLight~\cite{opendaylight}. Prefix de-aggregation is effective for hijacks of IP address prefixes larger than /24, but it might not work for /24 prefixes, as BGP advertisements of prefixes smaller than /24 are filtered by some ISPs.

In parallel to the mitigation, a monitoring service is running to provide real-time information about the mitigation process. This service uses again data from Periscope, RIPE RIS, and BGPmon to monitor/visualize the mitigation.

%% file: Experiments.tex
To evaluate ARTEMIS
, we conduct prefix hijackings against our own prefixes in the Internet. We use the PEERING testbed~\cite{Schlinker-PEERING-HotNets-2014}, which owns actual AS numbers (ASNs) and IP prefixes, and is connected to the Internet at multiple sites (university networks and IXPs)
. Through PEERING, we run a virtual AS, which announces a prefix and uses ARTEMIS to detect and mitigate hijackings for this prefix. We then announce the same prefix from another virtual AS of PEERING, emulating effectively a prefix hijacking attack. We associate different~sites with the two ASes, and denote them as \texttt{ASN-1} and  \texttt{ASN-2}. Each experiment consists of the following phases. 
 
\textbf{(Phase-1) Setup.} We announce an IP prefix, say {10.0.0.0/23}, from the legitimate owner of the prefix (\texttt{ASN-1}), and wait until the announcement becomes visible to all the LGs in our arsenal, i.e., for BGP convergence. 

\textbf{(Phase-2) Hijacking and Detection.} Then, from a different site of PEERING, \texttt{ASN-2} hijacks the prefix {10.0.0.0/23}, announcing it with \texttt{ASN-2} as the origin AS number. The new announcement disseminates in the Internet as well, and the ASes that are "closer", change their preferred path for the prefix to \texttt{ASN-2}. We measure the time until ARTEMIS detects the prefix hijacking by observing an announcement with an illegitimate origin AS in the data it processes from Periscope, RIPE RIS, and BGPmon.

\textbf{(Phase-3) Mitigation.} Immediately after the detection, ARTEMIS triggers prefix de-aggregation to mitigate the attack: it splits the hijacked prefix {10.0.0.0/23} into two more specific sub-prefixes, i.e., {10.0.0.0/24} and {10.0.1.0/24}, and announces them. The announcements for the /24 sub-prefixes disseminate in the Internet, and the routes change back to \texttt{ASN-1}, since the more specific /24 prefixes are preferred over the initial /23 prefix. We measure the time from the moment prefix de-aggregation is triggered until all the vantage points in our data have switched to the legitimate \texttt{ASN-1}. 

Our preliminary results over a few dozen experiments show that ARTEMIS needs (on average) {$45$secs} to detect the hijacking, {$15$secs} to announce the de-aggregated /24 prefixes (through the controller), and, after that, the mitigation is completed within {$5$mins}. In total, the hijacking is completely mitigated around {$6$mins} after it has been launched (which is smaller than the duration of {$>80\%$} of the hijacking cases observed in~\cite{Shi-Argus-IMC-2012}). The detection is faster because it needs \textit{at least one} observation of the bogus route, while the mitigation is completed when \textit{every router} has the legitimate route.


%% file: Demo.tex
The goal of the demo is to show that it possible to detect and mitigate BGP prefix hijackings in near real-time on the actual Internet. We will use ARTEMIS over the PEERING testbed to perform hijacking experiments, like in Section~\ref{sec:experiments}. 
Using the monitoring service of ARTEMIS, we will visualize in real-time how the hijacking incident propagates in the Internet, turning affected networks into the illegitimate AS. This, as well as the effect of the mitigation, will be demonstrated with a geographical visualization of vantage points around the globe that select the (il-)legitimate origin-AS. 